\documentclass[aps,prx,amsmath,amssymb,longbibliography,lengthcheck,showpacs,superscriptaddress,floatfix]{revtex4-1}
\usepackage{graphicx}
\usepackage{color}
\begin{document}
\title{Intermittent rearrangements accompanying thermal fluctuations distinguish glasses from crystals}
\author{Hideyuki Mizuno}
\email{hideyuki.mizuno@phys.c.u-tokyo.ac.jp} 
\affiliation{Graduate School of Arts and Sciences, The University of Tokyo, Tokyo 153-8902, Japan}
\author{Hua Tong}
\affiliation{Department of Fundamental Engineering, Institute of Industrial Science, The University of Tokyo, Tokyo 153-8505, Japan}
\affiliation{School of Physics and Astronomy, Shanghai Jiao Tong University, Shanghai 200240, China}
\author{Atsushi Ikeda}
\affiliation{Graduate School of Arts and Sciences, The University of Tokyo, Tokyo 153-8902, Japan}
\affiliation{Research Center for Complex Systems Biology, Universal Biology Institute, The University of Tokyo, Tokyo 153-8902, Japan}
\author{Stefano Mossa}
\email{stefano.mossa@cea.fr}
\affiliation{Univ. Grenoble Alpes, CEA, IRIG-MEM, 38000 Grenoble, France}
\affiliation{Institut Laue-Langevin, BP 156, F-38042 Grenoble Cedex 9, France}
\date{\today}
\begin{abstract}
It is a persistent problem in condensed matter physics that glasses exhibit vibrational and thermal properties that are markedly different from those of crystals. While recent works have advanced our understanding of vibrational excitations in glasses at the harmonic approximation limit, efforts in understanding finite-temperature anharmonic processes have been limited. It is well known that phonons in crystals couple through phonon-phonon interactions, an extremely efficient mechanism for anharmonic decay that is also important in glasses. Here, however, we show that an additional anharmonic channel of different origin emerges in the case of glasses, which induces intermittent rearrangements of particles. We have found that thermal vibrations in glasses trigger transitions among numerous different local minima of the energy landscape, which, however, are located within the same wide (meta)basin. These processes generate motions that are different from both diffusive and out-of-equilibrium aging dynamics. We suggest that the intermittent rearrangements accompanying thermal fluctuations are crucial features distinguishing glasses from crystals.
\end{abstract}
\maketitle
\section{Introduction}
\label{sect:intro}
The low-temperature thermal properties of glasses are markedly different from those of crystals, which is a persistent fundamental problem in condensed matter physics. For instance, as observed in many experiments (see, among others,~\cite{Zeller_1971,Phillips_1981,Graebner_1986}), the specific heat in glasses shows a linear temperature dependence at low temperatures, $C \propto T$, which is different from the Debye prediction for crystals, $C \propto T^3$~\cite{Ashcroft_1976,Kittel_1996}. A similar anomaly is observed in the heat conductivity, where $\kappa \propto T^2$ is different from $\kappa \propto T^3$ predicted for crystals. Because thermal behavior is determined by vibrational entities, these peculiar features have been rationalized at cryogenic temperatures by resorting to quite elusive localized excitations, the two-level systems~\cite{Anderson_1972,Phillips_1987,Galperin_1989}, whose actual existence and true nature are not yet fully understood~\cite{Heuer_2008,Jug_2016,Bonfanti_2017,Khomenko_2019}.

Recent advances have established a satisfactory understanding of the vibrational modes (eigenmodes) in glasses in the {\em harmonic} approximation. In this view, atoms undergo infinitesimal displacements around a stable configuration, the inherent structure, corresponding to a minimum of the associated potential energy landscape (PEL). It has been shown that quasi-localized vibrational~(QLV) modes emerge in the low-frequency part of the spectrum~\cite{Lerner_2016,Mizuno_2017,Shimada_2018,Wang_2019}, in addition to the phonon-like (acoustic) modes. At higher frequencies, however, disordered and extended vibrational modes are present~\cite{Wyart_2006,Silbert_2009,Mizuno_2017}, which are observed as excess modes over the Debye prediction for the vibrational density of states~(vDOS), the so-called Boson peak~\cite{buchenau_1984,Yamamuro_1996,Kabeya_2016}. The manner in which {\em anharmonicities} modify the above situation, possibly removing the system from the inherent structure, and the details regarding the role that they play in determining thermal properties are currently at the center of a lively debate.

In crystals at finite temperatures, phonons~\cite{Ashcroft_1976,Kittel_1996} excited by thermal fluctuations couple through phonon-phonon interactions~\cite{Ladd_1986,McGaughey_2004,McGaughey}, a mechanism well described by perturbation theories~\cite{Cowley_1963,Maris_1971}. In three-phonon interactions, one phonon splits into two different ones, or conversely, two phonons combine into a single excitation. In all cases, both energy and momentum are conserved.

Similarly, in glasses, vibrational eigenmodes also show anharmonic decay due to mode-mode interactions~\cite{Fabian_1996,Bickham_1998,Fabian_2003,Mizuno_2016}. In contrast to crystals, however, it has been suggested that a different kind of anharmonicity can emerge, inducing additional rearrangements of particles~\cite{Xu_2010}. Recently, we have shown that these rearrangements can be triggered by excitations of the lower-frequency modes, regardless of their phonon-like or QLV nature~\cite{Mizuno_2020}. Several experiments~\cite{Ruta_2012,Ruta_2014,Luo_2017} have indeed suggested the occurrence of fast dynamics of atoms in deep glass states, distinct from out-of-equilibrium aging dynamics. Numerical simulations~\cite{Ozawa_2015,Ozawa_2018} have also highlighted the presence of localized rearrangements in randomly pinned systems. All of the aforementioned works point to the existence of an anharmonic channel that induces additional particle rearrangements in disordered systems, which is absent in the corresponding crystals.

Here, we address the issue of how these anharmonic processes arise in glasses at finite temperatures by performing extensive molecular dynamics (MD) simulation of model atomic systems well below the glass transition temperature. First, we fully characterized the anharmonic decay due to the mode-mode interactions in terms of the lifetime of vibrational modes. We show that the numerical results can be closely reproduced by a theory with a third-order perturbation term, demonstrating that decaying processes involving three modes are dominant over other many-body processes. Second, we demonstrate unambiguously that thermal vibrations are accompanied by additional rearrangements of particles, corresponding to intermittent transitions between numerous local PEL minima that reside within one wide (meta)basin. Our results support the view that vibrational motions in glasses at nonzero temperatures follow the rugged profile of the underlying energy landscape, a situation that is markedly different from that observed in crystals.

In contrast to crystals, which are formed through a genuine thermodynamic transition from the liquid state, glasses form from the freezing of liquids because the dynamics become exceedingly slow on experimental timescales. In this sense, liquid-like features must remain in the glass. What we show here is that, indeed, remnants of the liquid state survive even far below the glass transition temperature in the form of unconventional dynamics superimposed upon thermal vibrations.
\section{Methods}
\label{sect:methods}
\subsection{Model and MD simulation}
\label{subsect:model}
We employ  a numerical atomistic model in three dimensions, which we have extensively studied in previous works~\cite{Mizuno2_2013,Mizuno_2014,Mizuno_2016}. Two point-like particles, $i$ and $j$, of type $\alpha$ and $\beta$, respectively, interact through a $12$-inverse power-law potential,
\begin{equation}
\frac{v_{\alpha \beta}(r)}{\epsilon} = \left( \frac{\sigma_{\alpha \beta}}{r} \right)^{12} + \left( \frac{\sigma_{\alpha \beta}}{r_c} \right)^{12}\left[ 12 \left(\frac{r}{r_c}\right)- 13 \right],
\label{eq:potential}
\end{equation}
where $r=r_{ij}$ is the particle distance and $\sigma_{\alpha\beta} = (\sigma_\alpha + \sigma_\beta)/2$, with $\sigma_{\alpha,\beta}$ as the particle diameters. The potential cut-off is $r_c=r_{c,\alpha\beta}=2.5\sigma_{\alpha\beta}$, and the form of Eq.~(\ref{eq:potential}) ensures that both potential and derived forces vanish continuously at $r=r_c$, avoiding artificial anharmonicities induced by discontinuities at $r_c$~\cite{Mizuno2_2016}. The mass, $m$, is identical for all particles.

We have considered both glassy (amorphous solid) and crystalline (completely ordered) states for comparison. For the latter cases, the particle diameter is $\sigma$ for all particles (monodisperse systems). In the glassy cases, to avoid crystallization, the samples are bidispersed, with $\sigma_1/\sigma_2 = 0.7$ and equivalent composition of two species, $x_{1,2}=N_{1,2}/N=1/2$ (with $N=N_1+N_2$ the total number of particles), where we fix an effective diameter $\sigma_\text{eff} \equiv (\sum_{\alpha,\beta=1,2}x_\alpha x_\beta \sigma^3_{\alpha\beta})^{1/3}$~\cite{bernu_1987} as $\sigma_\text{eff} = \sigma$. Additionally, we choose a (number) density $\hat{\rho}=N/V=1.015$, with the volume $V$ of the cubic simulation box. In the following, we employ $\sigma_\text{eff}=\sigma$, $m$, and $\epsilon$ as the units of distance, mass, and energy, respectively. Temperatures $T$ and frequencies $\omega$ are expressed in units of $\epsilon/k_B$~($k_B$ is the Boltzmann constant) and $\sqrt{\epsilon/(m\sigma^2)}$, respectively. The melting and glass-transition temperatures at the considered density are $T_m \simeq 0.6$ and $T_g \simeq 0.2$, respectively~\cite{bernu_1987}.

Crystalline phases are prepared by building a face-centered-cubic (FCC) perfect crystal, followed by equilibration in the $(NVT)$-ensemble, at the target value of $T$. For the glassy phases, we first equilibrate the system in the normal liquid state at $T=1.0$. We then instantaneously quench the system to the target temperature $T$ and equilibrate in the achieved (metastable) state by performing an extended $(NVT)$-ensemble run. In both cases, following the above equilibration procedures, we switch to the $(NVE)$-ensemble, regularly dumping the atomic coordinates, $\mathbf{r}(t) \equiv \{\mathbf{r}_i(t)\}_{i=1,\ldots, N}$. We have used the velocity Verlet algorithm for the numerical integration of the equations of motion with a time step $\delta t=5\times 10^{-3}$. We have performed all simulations by using the high-performance-computing MD tool LAMMPS~\cite{Plimpton_1995}.
\subsection{Trajectories and inherent structures}
\label{subsect:trajectories}
From the trajectory $\mathbf{r}(t)$, we can extract the time evolution of important thermodynamic quantities, including temperature and potential energy, 
\begin{equation}
T(t) \equiv \frac{1}{3N} \sum_{i=1}^N \left[ \frac{d \mathbf{r}_i(t)}{dt} \right]^2, \qquad \Phi(t) \equiv \sum_{i<j} v_{ij} \left( r_{ij}(t) \right),
\label{thermod12}
\end{equation}
where $r_{ij}(t) = \left|\mathbf{r}_i(t)-\mathbf{r}_j(t)\right|$, as well as the mean-squared displacements~(MSD),
\begin{equation}
\left<\left|\mathbf{r}_i(t) - \mathbf{r}_i(0) \right|^2 \right>_0 \equiv \left< \frac{1}{N} \sum_{i=1}^N \left| \mathbf{r}_i(t) -\mathbf{r}_i(0) \right|^2 \right>_0,
\label{msd}
\end{equation}
which indicates with $\left< \right>_0$ the ensemble average over the initial time $t=0$.

From $\mathbf{r}(t)$, which corresponds to $T>0$ system configurations, we can also extract the time series of the closest PEL minima ($T=0$) configurations, the inherent structures $\mathbf{r}_\text{IS}(t) \equiv \{\mathbf{r}_{\text{IS},i}(t)\}_{i=1,\ldots,N}$, by minimizing the potential energy of the instantaneous configurations at time $t$. We have employed the fast inertial relaxation engine (FIRE) minimization algorithm~\cite{Bitzek_2006}, but we have verified that other choices, such as the steepest descent method~\cite{Press_2007}, provide analogous results. From $\mathbf{r}_\text{IS}(t)$, we have calculated the potential energy in the inherent structure~\cite{Ozawa_2018},
\begin{equation}~\label{eqphiis}
\Phi_\text{IS}(t) \equiv \sum_{i<j} v_{ij} \left( r_{\text{IS},ij}(t) \right),
\end{equation}
where $r_{\text{IS},ij}(t) = \left| \mathbf{r}_{\text{IS},i}(t) - \mathbf{r}_{\text{IS},j}(t) \right|$.
\subsection{Particle rearrangements}
\label{subsect:rearrengements}
At each time $t$, we have also calculated the variations in $\mathbf{r}_\text{IS}(t)$ and $\Phi_\text{IS}(t)$ during a time lag $\Delta t = 10^{-1}$,
\begin{equation}
\begin{aligned}
\left| \Delta \mathbf{r}_\text{IS}(t) \right| & \equiv \left| \mathbf{r}_\text{IS}(t) - \mathbf{r}_\text{IS}(t-\Delta t) \right|, \\
& =\left( \sum_{i=1}^N \left| \mathbf{r}_{\text{IS},i}(t) - \mathbf{r}_{\text{IS},i}(t-\Delta t) \right|^2\right)^{1/2},
\label{eqdris}
\end{aligned}
\end{equation}
and
\begin{equation}~\label{eqdphiis}
\begin{aligned}
\left| \Delta \Phi_\text{IS}(t) \right| & \equiv \left| \Phi_\text{IS}(t) - \Phi_\text{IS}(t-\Delta t) \right|.
\end{aligned}
\end{equation}
Note that if $\left| \Delta \mathbf{r}_\text{IS}(t) \right|$ and $\left| \Delta \Phi_\text{IS}(t) \right|$ vanish, then no inherent-structure transition has occurred during the time lag. This outcome is always the case for crystals where, at the investigated values of $T$, the inherent structure continuously corresponds to the lattice site positions~\cite{Ashcroft_1976,Kittel_1996}. In contrast, if finite values are assumed, then a rearrangement of particles has taken place between $t-\Delta t$ and $t$.

In the following we show that rearrangements indeed occur in glasses and that the quantities $\{\Delta \mathbf{r}_{\text{IS},i}\}_{i=1,\ldots,N} \equiv \{\mathbf{r}_{\text{IS},i}(t) - \mathbf{r}_{\text{IS},i}(t-\Delta t)\}_{i=1,\ldots,N}$ provide the space-dependent displacement field associated with the rearrangement. Under these conditions, one can also estimate the number of particles participating in the rearrangements as~\cite{Mizuno_2020}
\begin{equation}
N_\text{rearr} \equiv \left[ \sum_{i=1}^{N} \left( \Delta \mathbf{r}_{\text{IS},i} \cdot \Delta \mathbf{r}_{\text{IS},i} \right) \right]^2 \left[ \sum_{i=1}^{N} \left( \Delta \mathbf{r}_{\text{IS},i} \cdot \Delta \mathbf{r}_{\text{IS},i} \right)^2 \right]^{-1}.
\label{eq:number}
\end{equation}
Note that the ratio $N_\text{rearr}/N$ is analogous to the participation ratio, which provides useful information about the localized/extended nature of the eigenvectors of the Hessian matrix~\cite{Mizuno2_2013,Mizuno_2016}.
\subsection{Lifetime of the vibrational eigenmodes}
\label{subsect:lifetimes}
In the cases where the investigated system does not undergo spatial rearrangements during the simulation time, the inherent structure corresponds to a fixed (immutable) configuration. As a consequence, the eigenvalues and eigenvectors extracted by a normal-mode analysis of the structure will also be fixed, making it meaningful to investigate the anharmonic decay of the eigenmodes triggered by the mode-mode interactions. This possibility is, however, meaningless in cases where rearrangements occur, and both the inherent structure and the associated eigenmodes are therefore time dependent.

In the cases where rearrangements do not occur, we can characterize the anharmonic decay due to mode-mode interactions in terms of the relevant lifetime as follows. We first diagonalize the dynamical (Hessian) matrix corresponding to the fixed inherent structure coordinates $\mathbf{r}_\text{IS}$, extracting the set of eigenfrequencies $\omega_k$ together with the corresponding eigenvectors $\mathbf{e}^k \equiv \{\mathbf{e}^k_i\}_{i=1,\ldots,N}$. Here, the index $k$ denotes the mode number, with $k=1,2,\cdots,3N-3$, where the three translational modes are not considered.

We next measure the lifetime $\tau_k$ of each mode $k$ by following~\cite{Mizuno_2016} and evaluating the time series of the vibrational energies associated with eigenmode $k$,
\begin{equation}~\label{energy}
\begin{aligned}
E_k(t) = \frac{1}{2}\omega_k^2 {A_k(t)}^2 + \frac{1}{2} \left[ \frac{d A_k(t)}{dt} \right]^2.
\end{aligned}
\end{equation}
Here, $A_k(t) = \left[ \mathbf{r}(t)-\mathbf{r}_\text{IS} \right] \cdot \mathbf{e}^k$ is the projection of the displacement vector from the inherent structure along the eigenvector of mode $k$, i.e., the vibrational amplitude along the mode $k$~\cite{Mizuno2_2016}. The first and second terms in Eq.~(\ref{energy}) correspond to the potential and kinetic energies, respectively. Due to the equipartition theorem, the time average $\left< E_k(t) \right>$ coincides with $T$~(see Fig.~S5 in the Supporting Information~(SI)). The normalized time correlation function of the energy fluctuations, $\delta E_k(t) = E_k(t) - \left< E_k(t) \right>$, is then calculated as
\begin{equation}
C_k(t) = \frac{\left< \delta E_k(t) \delta E_k(0)\right>_0}{\left< \delta {E_k(0)}^2 \right>_0}. 
\end{equation}
As demonstrated in~\cite{Mizuno_2016}~(see also Fig.~S6 in SI), $C_k(t)$ decays exponentially with time, and one can therefore extract the lifetime $\tau_k$ of eigenmode $k$ by imposing $C_k(t=\tau_k) = e^{-1}$~\cite{Ladd_1986,McGaughey_2004,McGaughey}.
\subsection{Theoretical prediction for the lifetimes}
\label{subsect:theory}
The above lifetime can be expressed as $\tau_k = 1/\Gamma_k$, where $\Gamma_k$ is the decay rate that can be evaluated analytically via a perturbation theory~\cite{Cowley_1963,Maris_1971,Fabian_1996,Fabian_2003}. By considering a third-order perturbation term, one obtains
\begin{equation}~\label{decayrate}
\begin{aligned}
\Gamma_k &= \frac{\hbar \pi}{4} \sum_{l=1}^{3N-3} \sum_{m=1}^{3N-3} \frac{\left| V_{klm} \right|^2}{\omega_k \omega_l \omega_m} \\
& \qquad \qquad \times \bigg[ \frac{1}{2}\left( 1+n_k+n_l \right)\delta \left( \omega_k-\omega_l-\omega_m \right) \\
& \qquad \qquad \qquad + \left( n_k-n_l \right) \delta \left( \omega_k+\omega_l-\omega_m \right) \bigg],
\end{aligned}
\end{equation}
where $n_k = \left[ \exp\left(\hbar \omega_k/k_B T \right) -1 \right]^{-1}$ is the Bose-Einstein occupation factor, $\delta(x)$ is the Dirac delta function, and $\hbar = h/2\pi$, with $h$ representing the Planck constant. $V_{klm}$ is related to the third-order derivative of the potential $\Phi_\text{IS}$ evaluated in the inherent structure and can be written as
\begin{equation}
\begin{aligned}
V_{klm} &= \sum_{i_1=1}^N \sum_{i_2=1}^N \sum_{i_3=1}^N \frac{\partial^3 \Phi_\text{IS}}{\partial \mathbf{r}_{i_1} \partial \mathbf{r}_{i_2} \partial \mathbf{r}_{i_3}} \mathbf{e}_{i_1}^k \mathbf{e}_{i_2}^l \mathbf{e}_{i_3}^m, \\
&= \sum_{i<j} A_{ij} \left( \mathbf{n}_{ij}\cdot \mathbf{e}^k_{ij} \right)\left( \mathbf{n}_{ij}\cdot \mathbf{e}^l_{ij} \right)\left( \mathbf{n}_{ij}\cdot \mathbf{e}^m_{ij} \right) \\
& \quad + \sum_{i<j} B_{ij}\bigg[ 
\left( \mathbf{e}^k_{ij}\cdot \mathbf{e}^l_{ij} \right)\left( \mathbf{n}_{ij}\cdot \mathbf{e}^m_{ij} \right) \\
& \qquad \qquad \qquad + \left( \mathbf{e}^l_{ij}\cdot \mathbf{e}^m_{ij} \right)\left( \mathbf{n}_{ij}\cdot \mathbf{e}^k_{ij} \right) \\
& \qquad \qquad \qquad + \left( \mathbf{e}^m_{ij}\cdot \mathbf{e}^k_{ij} \right)\left( \mathbf{n}_{ij}\cdot \mathbf{e}^l_{ij} \right)
\bigg],
\end{aligned}
\end{equation}
with
\begin{equation}
\begin{aligned}
A_{ij} &= \frac{1}{2} \frac{d^3 v_{ij}}{dr_{ij}^3} - \frac{3}{2r_{ij}} \frac{d^2 v_{ij}}{dr_{ij}^2} + \frac{3}{2r_{ij}^2} \frac{d v_{ij}}{dr_{ij}}, \\
B_{ij} &= \frac{1}{2r_{ij}} \frac{d^2 v_{ij}}{dr_{ij}^2} - \frac{1}{2r_{ij}^2} \frac{d v_{ij}}{dr_{ij}}.
\end{aligned}
\end{equation}
Here, $\mathbf{e}^k_{ij} \equiv \mathbf{e}^k_{i}-\mathbf{e}^k_{j}$, $\mathbf{n}_{ij} = (\mathbf{r}_i-\mathbf{r}_j)/\left| \mathbf{r}_i-\mathbf{r}_j \right|$ is the unit vector joining particles $i$ and $j$, and $\mathbf{r}_i$~($i=1,\ldots,N$) represents the atom positions in the inherent structure, $\mathbf{r}_{\text{IS},i}$.

In this work, we are interested in the classical limit that we can recover by considering $\hbar \rightarrow 0$ in Eq.~(\ref{decayrate}), thus obtaining
\begin{equation}~\label{decayrate2}
\begin{aligned}
\Gamma_k =& \frac{\pi k_B T}{4} \sum_{l=1}^{3N-3} \sum_{m=1}^{3N-3} \frac{\left| V_{klm} \right|^2}{{\omega_l}^2 {\omega_m}^2} \\
& \times \bigg[ \frac{1}{2}\delta \left( \omega_k-\omega_l-\omega_m \right) + \delta \left( \omega_k+\omega_l-\omega_m \right) \bigg].
\end{aligned}
\end{equation}
Note that the first term in the r.h.s. of Eqs.~(\ref{decayrate}) and~(\ref{decayrate2}) corresponds to a process where a mode $k$ splits into two modes $l$ and $m$, while the second term corresponds to the case where two modes $k$ and $l$ combine to create mode $m$.

Related to the expression of $\Gamma_k$, we also consider the joint two-mode density of the states~\cite{Fabian_1996,Fabian_2003},
\begin{equation}
\label{eq:joint}
\begin{aligned}
j(\omega_k) &= \frac{1}{(3N-3)^2}\sum_{l=1}^{3N-3} \sum_{m=1}^{3N-3} \\
& \qquad \bigg[ \frac{1}{2} \delta(\omega_k-\omega_l-\omega_m) + \delta(\omega_k+\omega_l-\omega_m) \bigg], \\
&= \int \frac{1}{2} g(\omega) g(\omega_k-\omega)d\omega + \int g(\omega) g(\omega_k + \omega) d\omega, \\
&\equiv j_\text{spl}(\omega_k) + j_\text{com}(\omega_k),
\end{aligned}
\end{equation}
where $g(\omega)$ is the vDOS~\cite{Mizuno2_2013,Mizuno_2016}. The term $j(\omega_k)$ accounts for the total number of combinations for the mode $k$ entering the three-mode couplings. The term $j_\text{spl}(\omega_k)$ describes processes where $k$ splits into two excitations, while $j_\text{com}(\omega_k)$ describes those where $k$ combines with another mode to create an additional mode. Note that $V_{klm}$ needs to be nonzero for the three-mode process to be active~(see Eqs.~(\ref{decayrate}) and~(\ref{decayrate2})). Since $g(\omega)$ assumes nonzero values only in the range $0 \le \omega \le \omega_\text{max}$, where $\omega_\text{max}$ is the maximum eigenfrequency, $j_\text{spl}(\omega_k)$ and $j_\text{com}(\omega_k)$ vanish for $\omega_k \to 0$ and $\omega_k \to \omega_\text{max}$, respectively. Combining and splitting processes thus appear to dominate at low and high frequencies, respectively, as shown in Fig.~\ref{fig4}.

\begin{table}[b]
\caption{\label{table1}
{MD simulation cases.}
We study glass and crystal of different system sizes $N$ and temperatures $T$. Simulation cases are denoted by ``Yes'' or ``No'', which indicate that rearrangements occur or do not occur, respectively, in the simulation time window.} 
\centering
\renewcommand{\arraystretch}{1.1}
\begin{tabular}{c|c|c|c|c|c|c}
\hline
\hline
& $T$ & $10^{-4}$ & $10^{-3}$ & $10^{-2}$ & $5\times 10^{-2}$ & $10^{-1}$ \\
\cline{1-7}
Glass & $N=4000$ & No & No & No & Yes & $ $ \\
& $32000$ & No & No & Yes & Yes & $ $ \\
& $256000$ & No & Yes & Yes & Yes & $ $ \\
\cline{1-7}
Crystal & $N=4000$ & $ $ & No & No & No & No \\
& $32000$ & $ $ & No & No & No & No \\
\hline
\hline
\end{tabular}
\end{table}

\begin{figure}[t]
\centering
\includegraphics[width=0.45\textwidth]{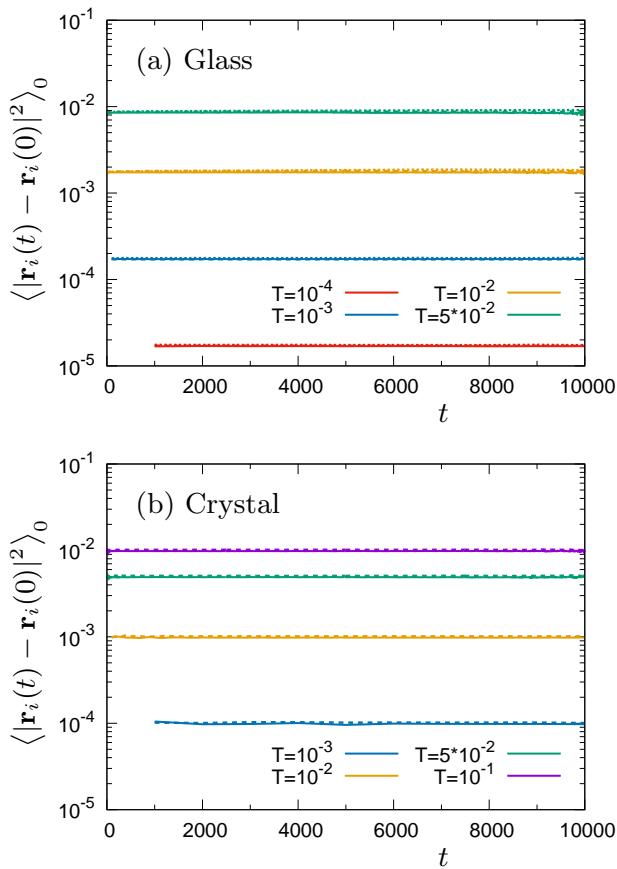}
\caption{\label{fig1}
{Mean-squared displacements.} MSD, $\left<\left| \mathbf{r}_i(t) - \mathbf{r}_i(0) \right|^2 \right>_0$, as a function of time $t$ for the glass (a) and the crystal (b) at the indicated values of $T$. We show data for $N=4000$~(solid lines) and $N=32000$~(dashed lines) in all cases and for $N=256000$~(dotted lines) for the glass. No relevant finite size effects are visible in these data.}
\end{figure}

\begin{figure}[t]
\centering
\includegraphics[width=0.45\textwidth]{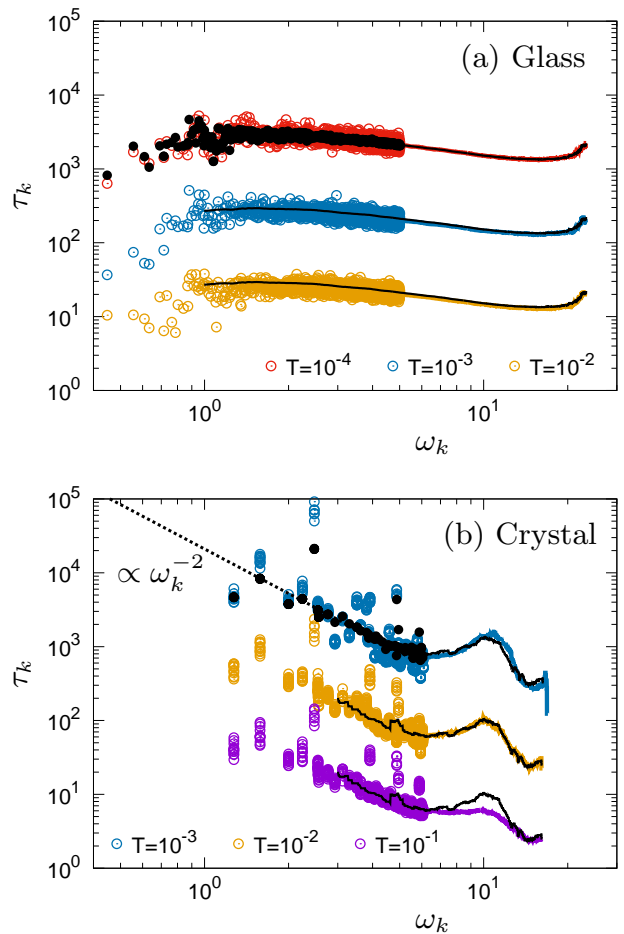}
\caption{\label{fig2}
{Lifetime of vibrational eigenmodes.} Lifetime $\tau_k$ as a function of eigenfrequency $\omega_k$ for the glass (a) and the crystal (b) at the indicated values of $T$, with $N=4000$. Open symbols represent the raw data for eigenmodes in the low-$\omega_k$ regime, while solid lines represent the values averaged over frequency bins of width $\delta \omega_k = 0.5$. In addition, black closed circles and black solid lines represent predictions of the perturbation theory of Eq.~(\ref{decayrate2}). In (b), we indicate with a dotted line the $\propto \omega_k^{-2}$ dependence expected in the low-$\omega_k$ regime. See also Fig.~S3 of the SI, where we plot data for $N=32000$.}
\end{figure}

\begin{figure}[t]
\centering
\includegraphics[width=0.45\textwidth]{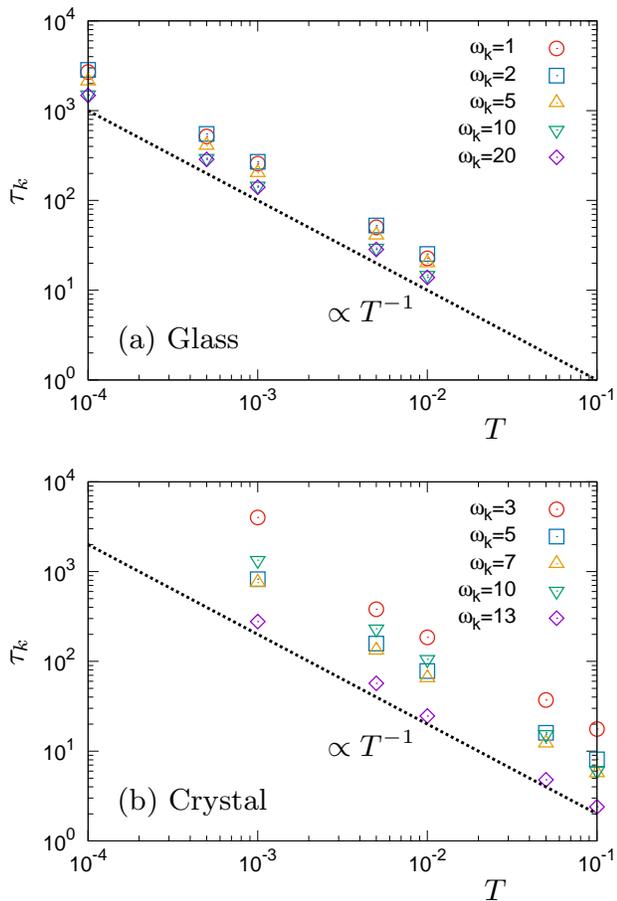}
\caption{\label{fig3}
{Temperature dependence of the lifetime of eigenmodes.} Lifetime $\tau_k$ as a function of temperature $T$, averaged over frequency bins of width $\delta \omega_k = 0.5$ centered at the indicated values of $\omega_k$ for the glass (a) and the crystal (b). The data are the same as those in Fig.~\ref{fig2}. The dotted line indicates the $\propto T^{-1}$ dependence, expected from the perturbation theory of Eq.~(\ref{decayrate2}).}
\end{figure}

\begin{figure}[t]
\centering
\includegraphics[width=0.45\textwidth]{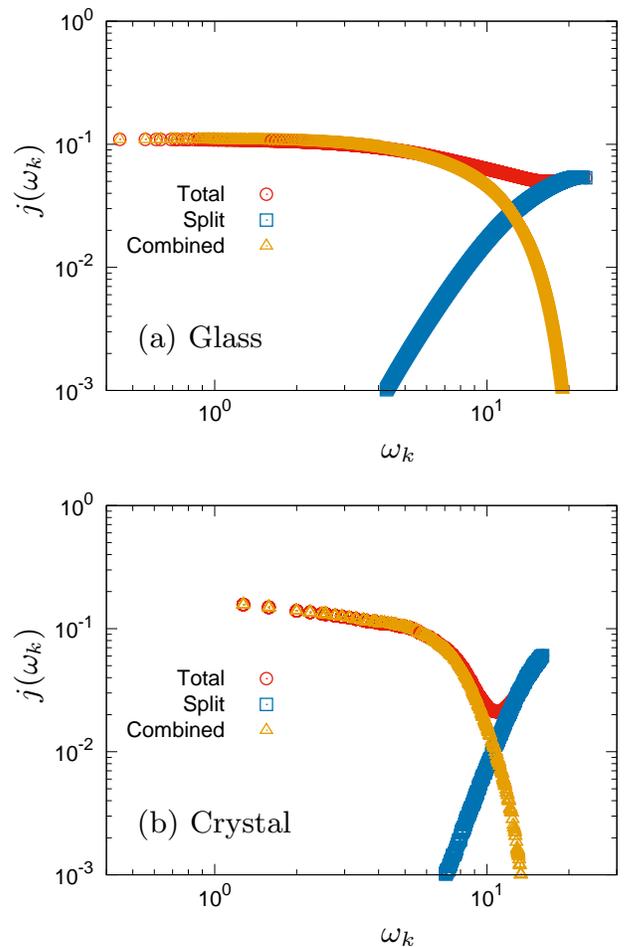}
\caption{\label{fig4}
{Joint two-mode density of states.} $j(\omega_k)$ as a function of the eigenfrequency $\omega_k$ for the glass (a) and the crystal (b), with $N=4000$. We plot the total density of the state as well as the first (split) and second (combined) terms of the r.h.s. of Eq.~(\ref{eq:joint}).}
\end{figure}

\begin{figure*}[t]
\centering
\includegraphics[width=0.99\textwidth]{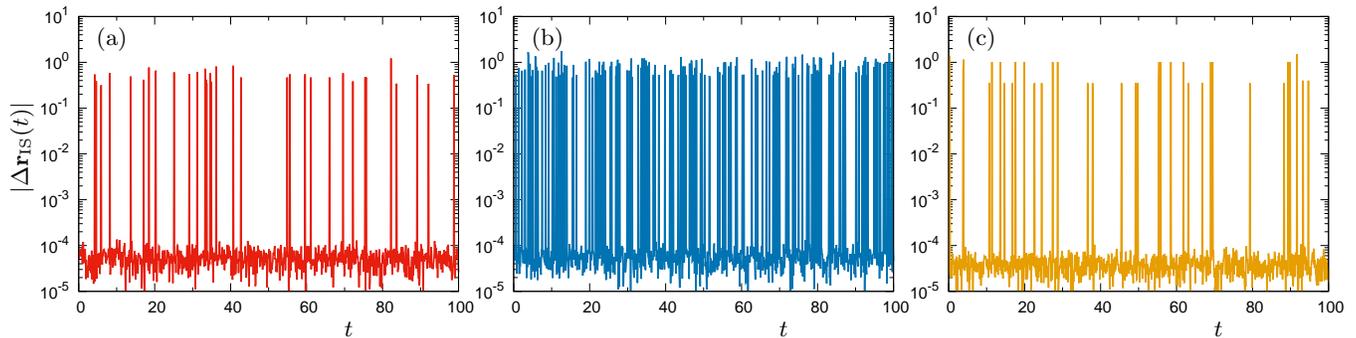}
\caption{\label{fig5}
{Intermittent particle rearrangements in the glass.} Variations in the inherent structure $\left| \Delta \mathbf{r}_\text{IS}(t) \right|$ during the time lag $\Delta t = 10^{-1}$ as a function of time $t$. Data are shown for the system size $N=32000$ and temperature $T=10^{-2}$ in (a), $N=32000$ and $T=5\times 10^{-2}$ in (b), and $N=4000$ and $T=5\times 10^{-2}$ in (c).}
\end{figure*}

\begin{figure}[t]
\centering
\includegraphics[width=0.45\textwidth]{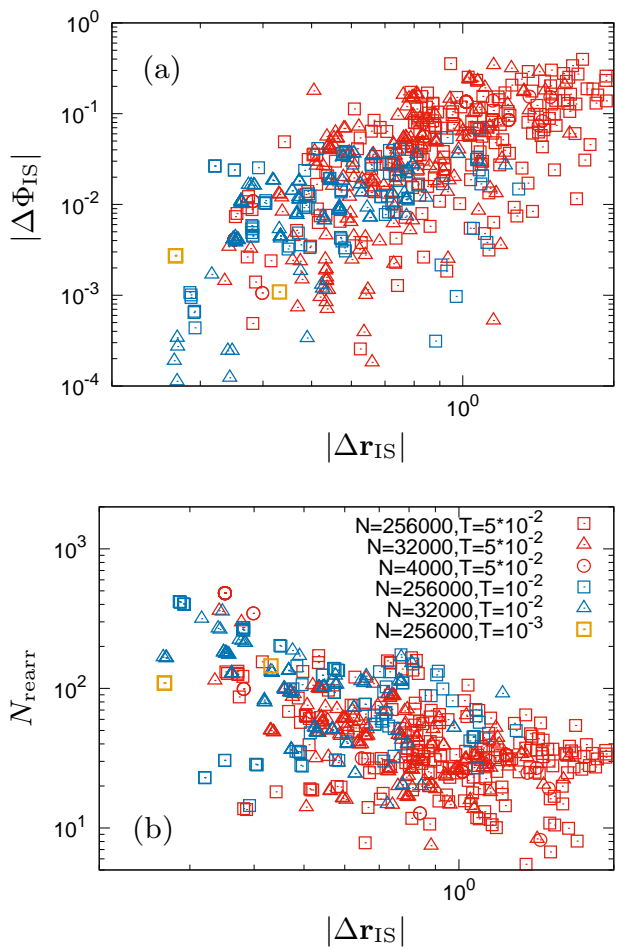}
\caption{\label{fig6}
{Variation in potential energy and number of rearranging particles in the glass.} Parametric plots of (a) variation in the potential energy, $\left| \Delta \Phi_\text{IS} \right|$, and (b) number of participating particles, $N_\text{rearr}$, versus $\left| \Delta \mathbf{r}_\text{IS} \right|$. We present the data together for the rearrangements that occur in our simulations at the indicated values of system size and temperature.}
\end{figure}

\begin{figure}[t]
\centering
\includegraphics[width=0.45\textwidth]{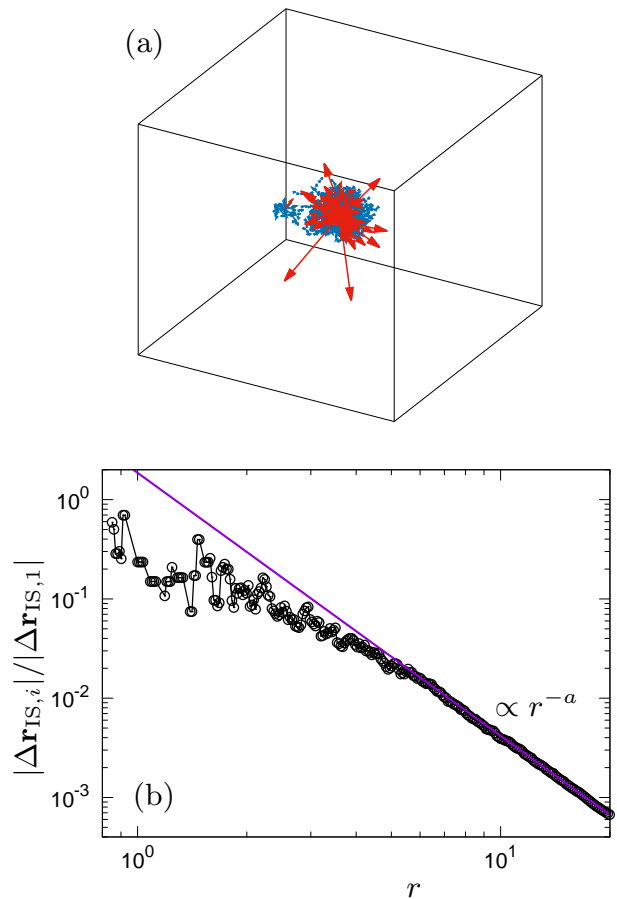}
\caption{\label{fig7}
{Quasi-localized displacement field of rearrangements in the glass.}
(a) Particle displacement $\Delta \mathbf{r}_{\text{IS},i}$ corresponding to the positions of particle $i$ in a representative inherent structure, $\mathbf{r}_{\text{IS},i}$. The $1000$ largest displacements rescaled as $\Delta \mathbf{r}_{\text{IS},i}\times 100$ are shown as arrows, and indicated in red are the largest $100$ among those. The system size and temperature are $N=256000$ and $T=10^{-2}$, respectively.
($\left| \Delta \mathbf{r}_\text{IS} \right| = 0.78$ and $N_\text{rearr} = 40$ for the present rearrangement.)
(b) Normalized displacement of particle $i$, $\left| \Delta \mathbf{r}_{\text{IS},i} \right|/ \left| \Delta \mathbf{r}_{\text{IS},1} \right|$, as a function of the distance $r$ from the particle $i=1$ with the largest displacement, $\left| \Delta \mathbf{r}_{\text{IS},1} \right|$. The tail at large distances decreases as a power law, $\left| \Delta \mathbf{r}_{\text{IS},i} \right| \propto r^{-a}$, with $a=2.7$ in this case. ($a$ has values of $2.5$ to $3$ over the course of our simulations.)}
\end{figure}

\begin{figure*}[t]
\centering
\includegraphics[width=0.99\textwidth]{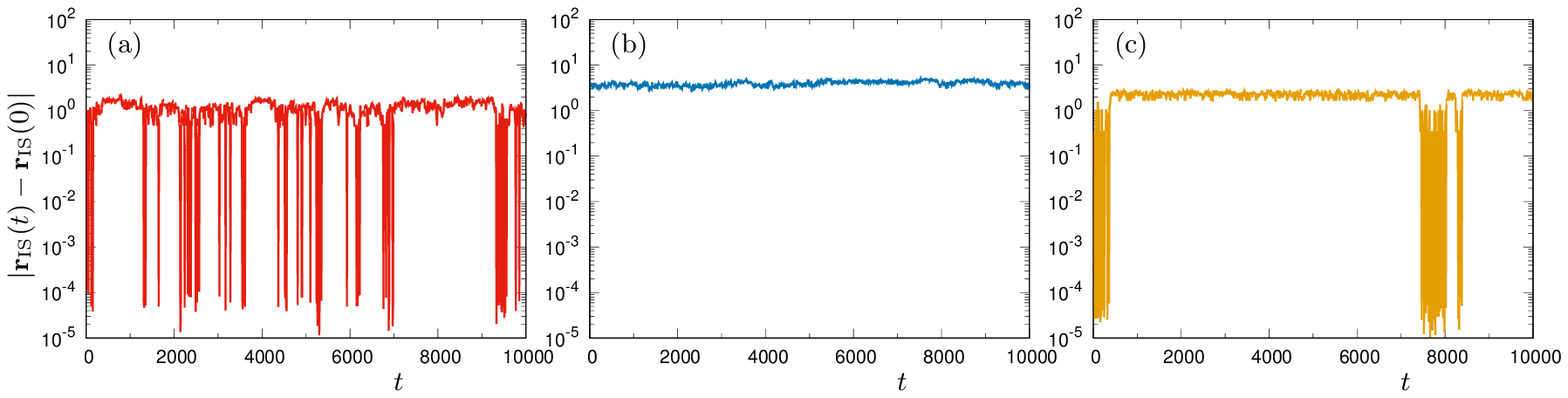}
\caption{\label{fig8}
{Time evolution of inherent structure in the glass.} The distance between $\mathbf{r}_\text{IS}(t)$ and $\mathbf{r}_\text{IS}(0)$, $\left| \mathbf{r}_\text{IS}(t) - \mathbf{r}_\text{IS}(0) \right|$, as a function of time $t$. The system size and temperature are $N=32000$ and $T=10^{-2}$ in (a), $N=32000$ and $T=5\times 10^{-2}$ in (b), and $N=4000$ and $T=5\times 10^{-2}$ in (c).}
\end{figure*}

\section{Results}
\label{sect:results}
In the following section, we discuss the MD simulation results for the glassy and crystalline states in three dimensions and the accompanying analyses~(see Sec.~\ref{sect:methods}). We report system details in Table~\ref{table1}. We have mainly studied two sample sizes, $N=4000$ and $32000$, for both glasses and crystals at temperatures ranging from $T=10^{-4}$ to $5\times 10^{-2}$ for the glass and $T=10^{-3}$ to $10^{-1}$ for the crystal, all well below the glass-transition~($T_g \simeq 0.2$) and melting~($T_m \simeq 0.6$) temperatures, respectively. A larger glassy instance, of size $N=256000$, has also been investigated to complement the analyses. For each investigated state, we have generated three independent system samples. Although data pertaining to one sample only are presented below, we have verified that our conclusions hold for the entire set of data.

We emphasize that in all cases of Table~\ref{table1}, the system evolves in a (meta)stable equilibrium state in the simulation time window. In Fig.~\ref{fig1}, we show the MSD of Eq.~(\ref{msd}) at the indicated values of $T$, which assumes a constant value in all cases, without any indication of time dependence. Note that at the same value of $T$, the MSD in the glass (a) is larger than that in the crystal (b), indicating that particle displacements are larger on average in the glass. The temperature $T(t)$ and potential energy $\Phi(t)$~(Eq.~(\ref{thermod12})) also are time independent, as we show in Figs.~S1 and~S2 of the SI. In particular, we do not observe any signal of diffusive or out-of-equilibrium aging dynamics in the glass, and can therefore conclude that the data are valid for extremely long-lived quasi-equilibrium states.

From the systems trajectories $\mathbf{r}(t)$ (at $T>0$), we extract the corresponding time series of the inherent structure, $\mathbf{r}_\text{IS}(t)$~(see Sec.~\ref{subsect:trajectories}). In crystals, the inherent structure is always identical, with particles vibrating around a perfect lattice structure~\cite{Ashcroft_1976,Kittel_1996}. In glasses, however, the inherent structure can change over time, and as a consequence, particles rearrangements can occur, as we discuss below. In Table~\ref{table1}, we use ``Yes'' for the cases where rearrangements were observed and ``No'' for the cases where rearrangements were not observed.

\subsection{Decay due to mode-mode interactions}
\label{subsect:decay}
We first characterize the anharmonic decay due to mode-mode interactions by focusing on the ``No'' cases of Table~\ref{table1} (with $N=4000$), where rearrangements never happen during the simulation time, and measuring the lifetime $\tau_k$ of each eigenmode $k$~\cite{Ladd_1986,McGaughey_2004,McGaughey,Mizuno_2016}~(see Sec.~\ref{subsect:lifetimes}). 

In Fig.~\ref{fig2}~(and Fig.~S3 of the SI), we show (open symbols) $\tau_k$ as a function of the corresponding eigenfrequency $\omega_k$ at the indicated values of $T$ for the glass (a) and the crystal (b). Glasses show a very mild $\omega_k$-dependence, whereas a much steeper variation is observed in crystals. These different features are partially related to the peculiar structures of the corresponding vDOS, as we discuss below. Temperature dependence is also worth noting, with $\tau_k \propto 1/T$ at fixed $\omega_k$ for both the glass and the crystal, as demonstrated directly in Fig.~\ref{fig3} at the indicated values of frequencies.

In addition, for the crystals shown in Fig.~\ref{fig2}(b), we observe $\tau_k \propto \omega_k^{-2}$ in the low-frequency regime, as expected in the continuum (hydrodynamic) limit~\cite{Maris_1971,Akhieser_1939}. (In Fig.~S3 of the SI, we plot additional sampling of the lower-frequency modes pertaining to the system with $N=32000$, where this behavior is even more evident.) Of course, we expect the continuum limit to also hold for the phonon-like vibrational modes in the low-frequency regime in our glasses~\cite{Mizuno_2017,Shimada_2018,Wang_2019}. A closer inspection of the data (Fig.~S3(a) of the SI) reveals that $\tau_k$ at low $\omega_k$ starts to separate into two branches, corresponding to phonon-like modes with higher $\tau_k$ and QLV modes with lower $\tau_k$. We need a more comprehensive investigation to confirm this conclusion.

The lifetimes can also be obtained as the inverse of the decay rate $\Gamma_k$, expressed analytically by means of the perturbation theory~\cite{Cowley_1963,Maris_1971,Fabian_1996,Fabian_2003} of Eqs.~(\ref{decayrate}) and~(\ref{decayrate2}), where we consider a third-order perturbation term. In Fig.~\ref{fig2}, we plot (with black circles and solid lines) the perturbation theory data $\tau_k = 1/\Gamma_k$ in the classical limit of Eq.~(\ref{decayrate2}), finding a very good agreement between the two sets of data. Note that the relationship $\Gamma_k \propto T$ in Eq.~(\ref{decayrate2}) directly implies $\tau_k \propto T^{-1}$, which is in agreement with the simulation data in Fig.~\ref{fig3}. Overall, these results demonstrate that three-mode couplings, including processes where one mode splits into two modes and processes where two single modes combine into one, are dominant over other many-mode processes, for both glasses and crystals. (Note that in the crystal at the highest considered $T=10^{-1}$, the simulation values are lower than the theoretical predictions, indicating that other many-mode processes are active in this high-temperature case.)

On this basis, we can understand the $T$- and $\omega_k$-dependencies of $\tau_k$ as follows. On one hand, the observation $\tau_k \propto 1/T$ originates from the Bose-Einstein occupation factor in the classical limit, $n_k \simeq k_B T/(\hbar\omega_k)$~\cite{Ashcroft_1976,Kittel_1996}. Therefore, as $T$ increases, the excitation of all eigenmodes grows, which increases the anharmonic decay and reduces their lifetimes. On the other hand, to investigate the eigenfrequency dependence, we plot the joint two-mode density of states $j(\omega_k)$~(Eq.~(\ref{eq:joint})) for the glass and the crystal in Figs.~\ref{fig4}(a) and~(b), respectively. If there are more (fewer) combinations that allow the mode $k$ to participate in the three-mode processes, its lifetime is obviously reduced (increased). This behavior is observed around $\omega_k \simeq 10$ in the crystal, where $j_k$ and $\tau_k$ are noticeably anti-correlated.

We cannot, however, fully explain the $\omega_k$-dependence of the anharmonic decay only in terms of the joint density of the states; the vertexes $V_{klm}$, which are particularly important for crystals, must also be considered~(see Eq.~(\ref{decayrate2})). In this case, $V_{klm}$ are nonzero only when the conservation of crystal momentum is satisfied by the involved modes $k$, $l$, and $m$, and vanish otherwise due to the periodicity of the crystalline lattice~\cite{Cowley_1963,Maris_1971}. Momentum conservation is thus necessary for the mode-mode interaction to activate in the crystal, which has an important implication. Indeed, from the data in Fig.~\ref{fig2}, we observe that at a given $T$, the $\tau_k$ of the crystal is greater overall (even an order of magnitude higher) than that of the glass. This observation is simply because the number of allowed three-mode combinations for the mode-mode interaction to become active in the crystal is lower than that in the glass, which is due to the much harder symmetry constraints imposed by the momentum conservation.

\subsection{Atomic rearrangements in glasses}
\label{sect.rearr}
We now turn to the cases where rearrangements of particles occur in the investigated time window, indicated in Table~\ref{table1} with ``Yes''. While rearrangements never take place in crystals, rearrangements of particles are observed in the glassy samples for all considered sizes, even at the very low indicated values of $T$. This effect is noticeably system-size dependent; the $N=256000$ system indicates rearrangements at $T=10^{-3}$, while $N=4000$ and $32000$ do not, and similarly, the $N=32000$ system indicates rearrangements at $T=10^{-2}$, while $N=4000$ does not.

This observation is in agreement with previous works~\cite{Xu_2010,Mizuno_2020} and can be explained by observing that larger values of $N$ (at constant $\hat\rho$) correspond with larger simulation box sizes, therefore including larger numbers of low-frequency modes that, in turn, trigger the observed rearrangements. As discussed in~\cite{Mizuno_2020}, we expect that in the $N\rightarrow \infty$ thermodynamic limit, rearrangements would manifest even at infinitesimally small temperature values. In the following section, we show data mainly pertaining to $N=32000$ and $4000$ and discuss the $T$- and $N$-dependence of the observed rearrangements in glasses. We have confirmed, however, that our conclusions are also valid for $N=256000$, and an additional detailed analysis of this case will be reported elsewhere.
\subsubsection{Intermittent rearrangements}
\label{subsubsect:intermittent}
In Fig.~\ref{fig5}, we show the data pertaining to the rearrangements in the glassy samples at $T=10^{-2}$ and $5\times 10^{-2}$, both approximately one order of magnitude lower than $T_g \simeq 0.2$. Specifically, we plot the time series of $\left| \Delta \mathbf{r}_\text{IS}(t) \right|$ defined in Eq.~(\ref{eqdris}), which is related to the eventual atomic position variations in the inherent structure $\mathbf{r}_\text{IS}(t)$ during a time lag of $\Delta t = 10^{-1}$. From these data, it is apparent that the rearrangements corresponding to the impulsive values (spikes) of $\left| \Delta \mathbf{r}_\text{IS}(t) \right|$ occur intermittently~\footnote{Values of $\left| \Delta \mathbf{r}_\text{IS}(t) \right| \sim 10^{-5}$ to $10^{-4}$ can be considered as corresponding to vanishing displacements within numerical precision, as demonstrated in the case of the crystal in Fig.~S4 of the SI.}.

When a rearrangement takes place, it is apparent that the potential energy $\Phi_\text{IS}(t)$ of Eq.~(\ref{eqphiis}) varies. In Fig.~\ref{fig6}(a), we show a parametric plot of $\left| \Delta \mathbf{r}_\text{IS}(t) \right|$ versus the corresponding $\left| \Delta \Phi_\text{IS}(t) \right|$ (see Eq.~(\ref{eqdphiis})) at the indicated values of $N$ and $T$, where each point corresponds to an observed rearrangement event. In addition to the observation that a larger rearrangement means a larger energy variation, as expected, two remarks regarding the $T$- and $N$-dependences are in order.

First, at the lower $T=10^{-2}$ and $N=32000$ in Fig.~\ref{fig5}(a), $\left| \Delta \mathbf{r}_\text{IS}(t) \right|$ can have values as small $10^{-1}$ (in units of the particle size $\sigma$). As already demonstrated in ~\cite{Mizuno_2020}~\footnote{Although in~\cite{Mizuno_2020} we considered a Lennard-Jones system, we confirm quantitatively similar results for the present system.}, this value corresponds to the minimum possible global displacement associated with the transition between adjacent inherent structures, while the transition paths to farther states are hindered on the short time scale $\Delta t=10^{-1}$. At the higher $T=5 \times 10^{-2}$ ($N=32000$, Fig.~\ref{fig5}(b)), in contrast, larger rearrangements with $\left| \Delta \mathbf{r}_\text{IS}(t) \right|$ of order $\sigma$ are also activated frequently. Note that the frequency of the rearrangements also increases with $T$, therefore substantially increasing the probability of long-range transitions. 

Second, comparing the data at the same $T$ but different system sizes $N=32000$ and $4000$ (in Figs.~\ref{fig5}(b) and (c), respectively), we observe that $\left| \Delta \mathbf{r}_\text{IS}(t) \right|$, which is {\em not} normalized by $N$, assumes very similar values in the two cases. The values for the $\left| \Delta \Phi_\text{IS}(t) \right|$ of Fig.~\ref{fig6}(a) are also very similar. Together, these observations imply a localized nature of the rearrangements, i.e., an $N$-independent fraction of atoms corresponds with the rearrangements, as we will discuss in the next section. We finally note that the frequency of the rearrangements is enhanced in larger systems due to the presence of a higher number of low-frequency modes~\cite{Mizuno_2017,Shimada_2018,Wang_2019}, as mentioned above.

In summary, our conclusion is that a temperature increase induces more frequent and larger rearrangements of particles, whereas an increase in the system size enhances the frequency of rearrangements but does not change their magnitudes (see also Fig.~\ref{fig6}(a)).

\subsection{Quasi-localized nature of the displacement field}
In Fig.~\ref{fig7}(a) we show an instance of the displacement field $\Delta \mathbf{r}_\text{IS}$ associated with a typical rearrangement (again with $\Delta t = 10^{-1}$). The localized nature of the field is clear, with only a limited fraction of the particles undergoing large displacements. More quantitatively, we can measure the number of particles involved in the rearrangement, $N_\text{rearr}$ (see Eq.~(\ref{eq:number})). In Fig.~\ref{fig6}(b), we show a parametric plot of $\left| \Delta \mathbf{r}_\text{IS} \right|$ versus $N_\text{rearr}$ for the rearrangements observed in our simulations. 

The data at different system sizes~(at the same temperature) overlap systematically, as is the case for $\left| \Delta \Phi_\text{IS}\right|$, which confirms a {\em strict} localization of the rearrangements with only tens to hundreds of particles involved~(also discussed in Ref.~\cite{Mizuno_2020}). Note that smaller rearrangements are less localized, now involving hundreds of particles, which is contrary to the observation of~\cite{Mizuno_2020}. This discrepancy may have arisen because the method we used to induce eigenmodes excitations was intrinsically different in~\cite{Mizuno_2020} than in this study. Indeed, in that work, only one target vibrational mode was excited selectively, while in the present work, all of the eigenmodes are equally excited by thermal fluctuations imposed by the coupling to the thermal bath with $T>0$. Further analysis is necessary to elucidate the mechanisms behind this disagreement.

The {\em quasi}-localized nature of the rearrangements can be seen in Fig.~\ref{fig7}(b), where we plot the displacement of particle $i$, $\left| \Delta \mathbf{r}_{\text{IS},i} \right|/ \left| \Delta \mathbf{r}_{\text{IS},1} \right|$, normalized to the observed maximum value $\left| \Delta \mathbf{r}_{\text{IS},1} \right|$ of particle $1$ as a function of the distance $r=r_{i1}$. The localized region is surrounded by a power-law tail, $\left| \Delta \mathbf{r}_{\text{IS},i} \right| \propto r^{-a}$, with an exponent $a\approx 2.5$ to $3$, distinct from the case of elastic deformation with $a=2$. Similar profiles for the displacements are observed in the elastic response to local forcing, as demonstrated in~\cite{Leonforte_2005,Lerner_2014}, and for QLV modes~\cite{Lerner_2016}, which, however, experience elastic-deformation behavior with $a=2$. We can attribute the steeper decay of the present far-field in the rearrangements to the nonlinear~\cite{Coulais_2014} or plastic~\cite{Maloney_2006} nature of the observed rearrangements.

We conclude with an observation. It was expected that correlations of the $\Delta \mathbf{r}_\text{IS}$ with the eigenmodes in the inherent structure $\mathbf{e}^k$. However, although some correlations with the low-frequency modes are found, they are generally very small. As pointed out in~\cite{Mizuno_2020}, excitations of the eigenmodes act as the trigger to induce the rearrangements, whereas the overall rearrangement patterns are determined by the rather involved global shape of the energy landscape~\cite{Middleton_2001,Reinisch_2004,Heuer_2008}, which emerges due to the extremely complex disorder-related, structural properties of amorphous systems~\cite{Hua_2018,Hua_2019,Tanaka_2019}.

\begin{figure*}[t]
\centering
\includegraphics[width=0.99\textwidth]{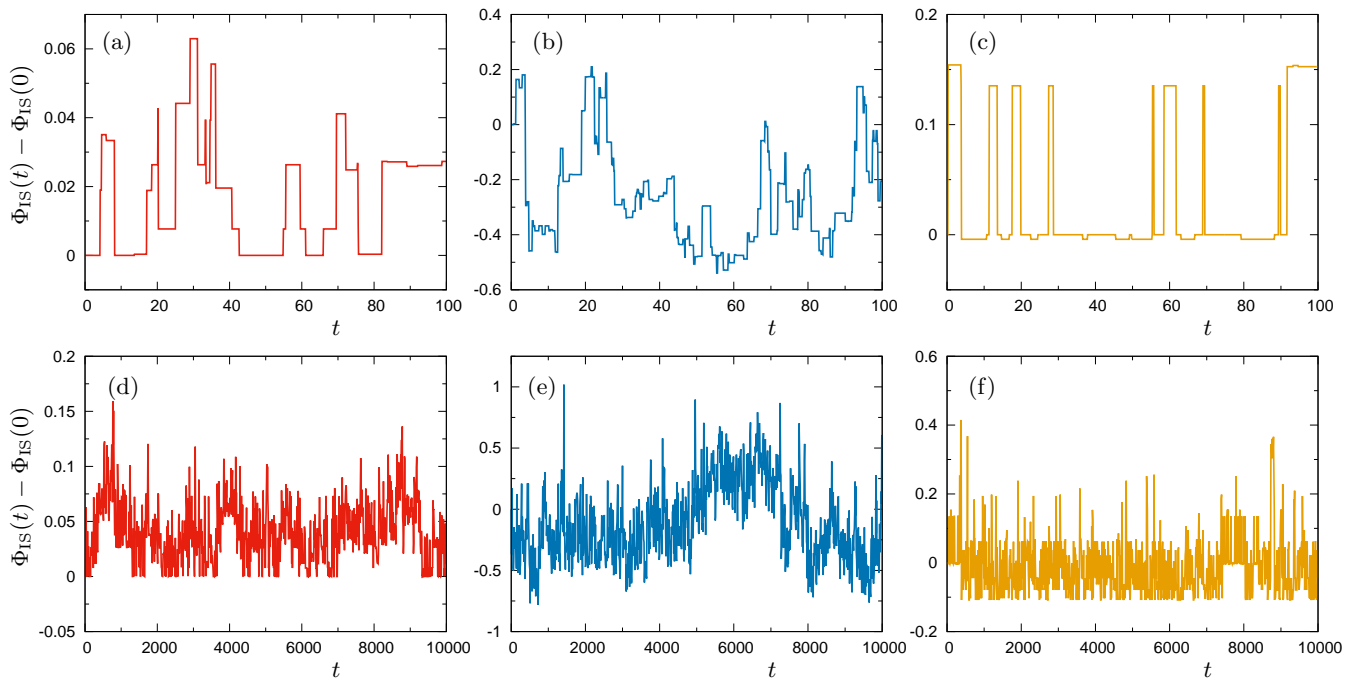}
\caption{\label{fig9}
{Time evolution of potential energy in the inherent structure of the glass.} The potential energy difference from the initial value, $\Phi_\text{IS}(t)-\Phi_\text{IS}(0)$, as a function of time $t$ during the shorter time scale~(a-c) and the longer time scale~(d-f). The system size and temperature are $N=32000$ and $T=10^{-2}$ in (a,d), $N=32000$ and $T=5\times 10^{-2}$ in (b,e), and $N=4000$ and $T=5\times 10^{-2}$ in (c,f).}
\end{figure*}

\subsubsection{Inherent structure transitions within one wide basin}
\label{susubsect:inherent}
We now focus on the long-time evolution of the inherent structures, $\mathbf{r}_\text{IS}(t)$, and discuss it in terms of the intermittent short-time rearrangement mechanism discussed above. In Fig.~\ref{fig8}, we show $\left| \mathbf{r}_\text{IS}(t)-\mathbf{r}_\text{IS}(0) \right|$ as a function of $t$ for the same three cases as in Fig.~\ref{fig5}. We immediately observe that $\left| \mathbf{r}_\text{IS}(t)-\mathbf{r}_\text{IS}(0) \right|$ spans a range of values that does not change during our simulation. This is of particular interest because it was expected that on long time scales, the system would be able to transition to increasingly more inherent structures, which would be signaled by an {\em increase} in $\left| \mathbf{r}_\text{IS}(t)-\mathbf{r}_\text{IS}(0) \right|$ with time. This observation therefore indicates that the system stays within one wide basin of the PEL~\cite{Middleton_2001,Reinisch_2004,Heuer_2008}.

We note that for the cases corresponding to a lower temperature or smaller system size (Figs.~\ref{fig8}(a) and~(c), respectively), the system evidently returns to the initial state, corresponding to a vanishing value of $\left| \mathbf{r}_\text{IS}(t)-\mathbf{r}_\text{IS}(0) \right|$, even if the intermittent character is highly variable. Conversely, in cases where $T$ or $N$ are increased (Fig.\ref{fig8}~(b)), the system never returns to the initial state, as indicated by the constantly finite values of $\left| \mathbf{r}_\text{IS}(t)-\mathbf{r}_\text{IS}(0) \right|$. This result is due to the larger thermal fluctuations at higher temperatures and more abundant low-frequency modes in larger systems, both of which induce more active accumulating rearrangements, as shown in Fig.~\ref{fig5}. 

The $\left| \mathbf{r}_\text{IS}(t)-\mathbf{r}_\text{IS}(0) \right|$ is a measure of the size of the explored basin. Since the system can explore a broader portion of the PEL at higher levels of thermal excitation, the size of the traversed basin increases with $T$, as is clear by comparing Figs.~\ref{fig8}(a) and~(b). We also recognize that the size of the basin increases with $N$ at constant $T$, as seen by comparing Figs.~\ref{fig8}(b) and~(c). We may expect $\left| \mathbf{r}_\text{IS}(t)-\mathbf{r}_\text{IS}(0) \right| \propto \sqrt{N}$ in the thermodynamic limit, which needs further investigation to be verified. Note that this square root dependence would only add a constant value to the MSD in all cases, thus generating no system size differences, as shown in Fig.~\ref{fig1}. The above observations suggest that in some conditions, the system explores a rugged energy landscape within the wide basin, as we discuss in the next section.

\subsubsection{Rugged profile of the potential energy landscape}
\label{susubsect:rugged}
In Fig.~\ref{fig9}, we plot the time evolution of the potential energy relative to the initial state, $\Delta\Phi_\text{IS}(t)=\Phi_\text{IS}(t) - \Phi_\text{IS}(0)$ (Eq.~(\ref{eqphiis})), for the shorter time scale~(up to $t=100$) in (a)-(c) and the longer time scale~(up to $t = 10000$) in (d)-(f). In all cases, the system undergoes transitions between numerous inherent structures with different values of $\Phi_\text{IS}(t)$, suggesting a significantly rugged profile of the explored PEL. Of course, at higher values of $T$, the system can explore a broader portion of the PEL due to thermal excitation, as can be seen by comparing panels (a) and (b) or (d) and (e).

Considering Fig.~\ref{fig9}, when comparing different system sizes ($N=32000$ in (b) and $N=4000$ in (c)) at constant $T$, fluctuations in the $\Phi_\text{IS}(t)$ clearly become more prominent at the larger $N$, indicating that larger systems have a more refined roughness structure. On the other hand, as shown in panels (e) and (f), the range that $\Phi_\text{IS}(t) - \Phi_\text{IS}(0)$ explores remains finite even in the long-time scale, which then becomes wider for the larger $N$. Since $\left| \Phi_\text{IS}(t) - \Phi_\text{IS}(0) \right|$ provides another indicator of the size of the explored basin, it was expected that $\left| \Phi_\text{IS}(t) - \Phi_\text{IS}(0) \right| \propto N$ in the $N\rightarrow \infty$ limit, which is consistent with $\left| \mathbf{r}_\text{IS}(t) - \mathbf{r}_\text{IS}(0) \right| \propto \sqrt{N}$. We therefore conclude that in the thermodynamic limit, the roughness of the energy landscape becomes infinitesimally small, for a specified wide metabasin.

Note that a mean-field theory based on the replica technique~\cite{simpleglass} predicts rugged~(or hierarchical) structure of the free energy landscapes in the {\em marginally stable} phase of glasses. This has been confirmed by recent numerical~\cite{Berthier_2016,Scalliet_2017,Scalliet_2019} and experimental~\cite{Hammond_2020} works, which have directly connected these PEL features to replicas. Different from these approaches, the present work focuses on the time evolution of the inherent structures and establishes a clear connection between the system rugged energy landscape and the intermittent rearrangements of the particles. Also note that the systems investigated here are far from the {\em jamming} regime and have localized rearrangements, which is consistent with numerical observations~\cite{Scalliet_2017}. We expect that the rearrangements become collective and extended if density is lowered towards the jamming regime~\cite{Berthier_2016,Scalliet_2019}. 

Finally, we observe that the marginally stable phase is also predicted theoretically~\cite{Ikeda2_2019} for disordered crystals~\cite{Mizuno2_2013,Mizuno_2014,Mizuno_2016,Tong_2015,Charbonneau_2019}, which are positionally ordered on the lattice structures but characterized by some form of disorder in the size of the particles (dispersion), or in the energy scales associated with the interparticle potential. It could be interesting to clarify whether intermittent rearrangements similar to those observed here for glasses also occur in those systems.
\section{Discussion and conclusions}
\label{sect:conclusion}
In this paper we have shown that, contrary to crystals, glasses at finite temperatures undergo not only the decay of the vibrational excitations due to mode-mode interactions {\em (I)} but also the rearrangement of particles {\em (II)} {\em in addition} to the expected anharmonic displacements. 

For crystals in all cases, or for glasses at very low temperatures, the representative point of the system on the PEL is trapped in a single inherent structure. In this situation, we have demonstrated that no additional atomic rearrangements can occur aside from the displacements associated with thermal fluctuations. We have also demonstrated that the excitations of the eigenmodes in glasses undergo anharmonic decay due to the mode-mode interactions {\em (I)}, which is analogous to the well-understood behavior of phonons in crystals. We have numerically measured the lifetime associated with the entire spectrum of eigenmodes and have shown that they are well reproduced by a third-order perturbation theory for both glasses and crystals. This result clearly demonstrates that three-mode processes, where one mode splits into two excitations or two modes combine into a single excitation, are dominant over other possible many-mode processes. In the absence of rearrangements, as expected, anharmonic processes are therefore common to glasses and crystals, as well as any solid-state material.

In contrast, in case {\em (II)}, we have demonstrated that rearrangements of particles, in addition to the anharmonic displacements, intermittently occur in glasses even at temperatures that are orders of magnitude lower than the glass-transition temperature. These rearrangements indicate the rugged profile of the underlying PEL, which is contrary to the case of crystals. We have also found that the additional rearrangements are quasi-localized in space. More specifically, tens to hundreds of particles undergo large displacements in a well-defined local region, surrounded by a far-field tail that is characterized by a power-law decay of displacements. Notably, the MSD and important thermodynamic quantities show no indications of diffusive or aging behavior, suggesting that intermittent rearrangements are distinct from these kinetic phenomena. The absence/presence of these additional entities is, as a consequence, a crucial distinction between ordered and disordered states of matter.

Remarkably, in the $N \to \infty$ thermodynamic limit, the particle rearrangements can be induced by infinitely small thermal fluctuations~\cite{Mizuno_2020}, as it was previously reported that infinitesimal mechanical strain can cause rearrangements of particles (i.e., plastic events)~\cite{Karmakar_2010}. These extremely fragile properties can be explained by the concept of {\em marginal stability}~\cite{Muller_2015}, which suggests that when a system is quenched from the liquid state, it freezes to the glassy solid-state as soon as it acquires stability, such that the frozen phase is exactly on the verge of instability. Marginal stability has been predicted by theories based on the replica technique~\cite{simpleglass} and the coherent potential or effective medium approximations~\cite{schirmacher_2006,schirmacher_2007,DeGiuli_2014,Shimada_2020}.

Experiments on structural glasses~\cite{Ruta_2012,Ruta_2014,Luo_2017} have detected the fast dynamics of atoms distinct from aging dynamics. It is natural to speculate that these fast dynamics are due to anharmonicity-related intermittent rearrangements similar to those discussed in this work. In addition, localized rearrangements have been observed in simulations of randomly pinned systems~\cite{Ozawa_2015,Ozawa_2018}. Due to pinning, these structures stay frozen even at temperatures orders of magnitude higher than those we considered here, but stronger thermal excitations lead to additional dynamics that are detected beyond the typical plateau in the MSD. Although the overall dynamics are very different in structural glasses and pinned systems, intermittent rearrangements and exploration of a rugged energy landscape therefore seem to be a common features of disordered systems, likely including disordered crystals~\cite{Mizuno2_2013,Mizuno_2014,Mizuno_2016,Tong_2015,Charbonneau_2019}. This is an interesting direction to investigate in the future.

Other open issues include the characterization of {\em effective} vibrational eigenmodes at finite temperatures. Indeed, although rearrangements occur intermittently in glasses, we have shown that these systems always stay within one wide metabasin of the underlying PEL. It is therefore reasonable to consider effective or coarse-grained vibrations, which would also comprise the effects of the rearrangements, a program that could be probably realized exploiting covariance matrix methods~\cite{Brito_2010,Henkes_2012}. Clarifying this point would be extremely useful in view of very recent work~\cite{Prasenjit_2019}, where an attempt was made to characterize effective eigenmodes as predictors of plastic instability at finite temperatures. 

Another interesting point is the impact of rearrangements on material properties. For instance, an analytical formulation for the heat conductivity based on the Green-Kubo formulation has been proposed~\cite{Allen_1993,Feldman_1993} and recently developed~\cite{Isaeva_2019,Simoncelli_2019}. Although these formulations can be applied to both glasses and crystals, they rely, however, on a quasi-harmonic approximation. They therefore incorporate mode-mode interactions but neglect the particle rearrangements discovered here. In addition, in~\cite{mizuno2019impact,mizuno2019impact2,Wang_2020}, it was found that sound damping in glasses varies with temperature as $\propto \sqrt{T}$, in contrast with the expected $\propto T$ behavior, a discrepancy that cannot be explained in terms of mode-mode interactions only. We speculate that the rearrangements observed here could indeed contribute to this anomalous temperature dependence. We also note that the theory of elastic heterogeneities~\cite{schirmacher_2006,schirmacher_2007} predicts the $\propto \sqrt{T}$ variation near the elastic instability~\cite{marruzzo2013vibrational,ferrante2013acoustic}. The effects of the rearrangements may therefore be implicitly included in the distributions of local elastic constants~\cite{Wagner_2011,Mizuno_2013}, the most important component of the theory. 

Finally, we mention that our simulations are classical and, therefore, do not include any potentially important quantum mechanisms that are known to be needed for a deep understanding of the anomalies in the low-temperature properties of glasses~\cite{Zeller_1971,Phillips_1981,Graebner_1986}. In the presence of quantum effects, vibrational states are populated according to the Bose-Einstein distribution, suggesting that lower (higher)-frequency modes are more (less) excited than what considered in classical calculations. More importantly, we expect that rearrangements can also be induced by the quantum tunneling processes~\cite{Jug_2016,Bonfanti_2017,Khomenko_2019}, which would establish a direct link with the two-level systems that have been used to explain some of the aforementioned anomalous glass properties~\cite{Anderson_1972,Phillips_1987,Galperin_1989}. All of these are open issues that should be addressed in the future.
\begin{acknowledgments}
H. M., H. T., A. I. are supported by JSPS KAKENHI Grant Numbers 17H04853, 18H05225, 18H03675, 19H01812, 19K14670, 20H01868, 20H00128, and Specially Promoted Research (25000002). S. M. is supported by ANR-18-CE30-0019 (HEATFLOW). This work has also been supported by the Asahi Glass Foundation.
\end{acknowledgments}
\bibliographystyle{apsrev4-1}
\bibliography{reference}
\end{document}